\documentclass[a4paper,fleqn]{cas-dc}

\usepackage[utf8]{inputenc}
\usepackage[numbers]{natbib}
\usepackage{amsmath, amssymb, bbm, mathtools, amsthm,bm}
\usepackage{graphicx}

\usepackage{caption}
\usepackage{subcaption}
\usepackage{enumitem}
\usepackage{algorithm,algorithmic}
\usepackage{nomencl}
\makenomenclature
\usepackage{tikz}
\usetikzlibrary{positioning,fit,backgrounds}
 
\usetikzlibrary{shapes,arrows}
\usepackage{mathptmx}

\begin{document}

\shorttitle{}

% Short author
\shortauthors{Preprint}

\title[mode=title]{Chance-constrained Solar PV Hosting Capacity Assessment for Distribution Grids Using Gaussian Process and Logit Learning %\vspace{-10mm}
}

\author[inst1]{Sel Ly}

\affiliation[inst1]{organization={School of Electrical and Electronics Engineering, Nanyang Technological University}, %Department and Organization
            %addressline={Nanyang Technological University}, 
            city={Singapore},
            %postcode={639798} 
            %state={State One},
            country={639798}
            }
            
\author[inst1]{Anshuman Singh}
\author[inst1]{Petr Vorobev}
\author[inst1]{Yeng Chai Soh}
\author[inst1]{Hung Dinh Nguyen}
\fnmark[1]
\fntext[1]{Corresponding Author}
\nonumnote{Authors are with School of Electrical and Electronics Engineering, Nanyang Technological University, Singapore: \textit{\{sel.ly, anshuman004, petr.vorobev,  hunghtd\}@ntu.edu.sg    } }

\begin{abstract}
Growing penetration of distributed generation such as solar PV can increase the risk of over-voltage in distribution grids, affecting network security. Therefore, assessment of the so-called, PV hosting capacity (HC) - the maximum amount of PV that a given grid can accommodate becomes an important practical problem. In this paper, we propose a novel chance-constrained HC estimation framework using Gaussian Process and Logit learning that can account for uncertainty and risk management. Also, we consider the assessment of HC under different voltage control strategies. Our results have demonstrated that the proposed models can achieve high accuracy levels of up to 93\% in predicting nodal over-voltage events on IEEE 33-bus and 123-bus test-cases. Thus, these models can be effectively employed to estimate the chance-constrained HC with various risk levels. Moreover, our proposed methods have simple forms and low computational costs of only a few seconds.
\end{abstract}


\maketitle

\nomenclature{$V_k, \theta_k$}{Voltage magnitude and phase at bus \textit{k}}
\nomenclature{$\underline{V},\overline{V}$}{Lower and upper limits on voltage magnitudes}
\nomenclature{$V_{max}$}{Maximum of nodal voltage magnitudes}
\nomenclature{$P_k,Q_k$ }{Active/reactive power injections at bus \textit{k} }
\nomenclature{$\theta_{kl}$}{Phase difference between buses $k$ and $l$ }
\nomenclature{$Y_{kl}$}{Element $(k,l)$ of admittance matrix Y} 
\nomenclature{$ \mathcal{L}, \mathcal{G}$}{Sets of all loads and generators in the power network}
\nomenclature{$N^b$ }{The number of buses in the power network.}
\nomenclature{$\overline{P}_{d}$}{Peak demand of the network (MW)}
\nomenclature{$P_{d,k}/ P_{g,k}$}{Power demand / Solar PV generation at bus $k$}
% \nomenclature{$P_{g,k}$}{Solar PV power generation at bus $k$}
\nomenclature{$P_{nd}/ P_{ng}$}{Normalized demand / Solar PV generation}
% \nomenclature{$P_{ng}$}{Normalized Solar PV power generation}
\nomenclature{$P_{L,k}$}{The net power injection (net load) at bus $k$}
\nomenclature{$X$}{Ratio of the total PV installation capacity to the system peak load (PV level)}
\nomenclature{$\mathcal{C}_k$}{The solar PV rated capacity installed at bus $k$}
\nomenclature{$\mathcal{\overline{C}}_k$}{Upper limit of the solar PV rated capacity installed at bus $k$}
\nomenclature{$PV^{loc}$}{ The set of solar PV locations (PV buses) }
\nomenclature{$s_{kl}$}{The apparent power flow between bus $k$ and $l$}
\nomenclature{$\underline{s}_{kl},\overline{s}_{kl} $}{Lower and upper limits of the apparent power flow}
\nomenclature{$\mathcal{B}$}{The set of lines (branches) in the power network}
\nomenclature{$S_{g,k}$}{ Apparent power generated by PV unit at bus $k$}
\nomenclature{$C_\rho$ }{Gaussian copula with its parameter $\rho$}
\nomenclature{$\Omega$}{The set of correlated scenarios between
solar irradiance and load demand }
\nomenclature{$\omega$}{Scenario from the uncertain set $\Omega$}
\nomenclature{$\beta\in (0,1)$}{Risk level for the over-voltage event}
\nomenclature{$R(x)$}{Predicted probability of the over-voltage event (risk index), given the PV level $X=x$ }
\nomenclature{$f(x)$}{The predictive GP model indexed by the PV level input $X=x$}
\nomenclature{$\mu_f(x^*), \sigma^2_f(x^*)$}{Predicted mean and variance values of $V_{max}$ by the GPR, given the unseen PV level $X=x^*$}
\nomenclature{$\bm{K}$}{Kernel matrix of the GP model}
\nomenclature{$\lambda,\tau$}{Hyperparameters of the GP kernel function}
\nomenclature{$\widetilde{V}$}{Binary variable indicating the over-voltage event}
\nomenclature{$\Phi$}{CDF of standard Gaussian distribution $\mathcal{N}(0,1)$}
\printnomenclature

\section*{Abbreviations}
Below in Table \ref{tab:abbv}, we provide a list of abbreviations/ terminologies used in the paper.

\begin{table}[!ht] 
    \centering
\caption{Abbreviations/terminologies.} \label{tab:abbv}
    \begin{tabular}{p{0.28\linewidth} | p{0.59\linewidth}}
   \textbf{Name} & \textbf{Meaning} \\
   \hline
   \textbf{CC} & Chance-constrained.  \\
   \textbf{CCO} & Chance-constrained optimization. \\
    \textbf{HC} & Hosting capacity. \\
    \textbf{GP} & Gaussian process. \\
    \textbf{GPR} & Gaussian process regression. \\
\textbf{LoR} & Logistic regression. \\
\textbf{CDF} & Cumulative distribution function. \\
\textbf{MCS} & Monte Carlo simulation.\\
\textbf{PV} & Photovoltaic.\\
% \textbf{WT} & Wind turbine.\\
 \textbf{RES} & Renewable energy source. \\
\textbf{DER} & Distributed energy resource. \\
     \textbf{GP-WoCC-HC} & Mean and bounds of HC estimations resulting from the GP learning without considering CCO.  \\
     \textbf{GP-CC-HC} &  HC estimations resulting from the GP learning and CCO.  \\
     \textbf{Logit-CC-HC} & HC estimations resulting from the logit learning and CCO.    \\
     \textbf{PCE-CC-POPF} & Polynomial chaos
expansion-based chance-constrained probabilistic optimal power flow. \\ 
   \textbf{Q control} & Droop-based reactive power control. \\
     \textbf{PF control} & Droop-based power factor control. \\
      \textbf{ESS control} & Droop-based energy storage system control. \\ \hline
    \end{tabular}
\end{table}

\color{black}

\section{Introduction}
% needs further improvement
Integrating renewable energy sources (RESs) such as solar PV  at the distribution network is a key element in the transition towards a more sustainable and resilient energy future~\cite{sinsel2020challenges}. 
However, a technical limit exists on the maximum amount of RESs that can be installed in a network. This limitation is primarily attributed to over-voltages caused by reverse power flows~\cite{weng2021fixed, eltigani2015challenges}. As RESs at the distribution level are non-dispatchable, these reverse power flows cannot be effectively managed or controlled.
% However, the variability and intermittency of RESs can lead to technical challenges such as over-voltage and/or over-current violations \cite{weng2021fixed, eltigani2015challenges}. 
Thus, to assess the impact of RESs on the network, it is essential to perform an estimation of the so-called, network's hosting capacity (HC), i.e. the maximum amount of distributed generation that can be connected to the network without violating its operational limits and without the need for network expansion~\cite{mulenga2021solar}.

Overestimation of HC can lead to grid instability, increased likelihood of power outages, and potential damage to equipment \cite{kharrazi2020assessment, ali2023calculating}. On the other hand, underestimation of HC can lead to the under-utilization of the grid capabilities to host RES, which can impact the economic viability of renewable energy projects and delay the transition towards a more sustainable energy future. Thus, it is essential to perform an accurate estimation of the network's HC for maintaining grid stability and ensuring safe, reliable operation.

Stochastic estimation of HC can be performed through Monte Carlo (MCS) simulation-based methods which include the generation of a large number of location-size scenarios for a given PV penetration level, followed by load flow analysis of each scenario. Frameworks to estimate HC by the MCS method have been proposed in works \cite{dubey2016estimation, arshad2019stochastic,ding2016distributed}. 

However, MCS-based methods only generate a small fraction of location-size scenarios out of all possible combinations. Hence, these methods need to be re-run multiple times to obtain confidence bounds on the HC value, resulting in thousands of MC simulations. Additionally, the above works do not consider the uncertainties of load and PV output, and only worst-case scenarios are considered, i.e. maximum RES output and minimum load obtained from historical data. This results in a conservative estimation of the network's HC.

% uncertain load-generation paper
Various methods have been proposed to address the uncertainty in load and generation for MCS-based HC estimation frameworks. Approaches include probabilistic modelling of demand and generation using various distributions \cite{ayaz2024probabilistic}, interval and affine arithmetic \cite{moro2024distributed}, possibilistic methods with membership functions \cite{yao2022possibilistic}, and Herman-Beta extended transform-based probabilistic power flow \cite{chihota2022stochastic}.  Authors in \cite{wang2019interval} have proposed a Monte Carlo-based approach for evaluating the PV hosting capacity of a three-phase unbalanced network. In this work, the maximum voltage for each location-size scenario is obtained and an interval distribution for this relationship is constructed. The HC is evaluated such that the probability of overvoltage is within a given interval. However, these studies do not account for the correlation between load and Distributed Energy Resource (DER) generation, which can impact reliability assessment, as highlighted in \cite{qin2013incorporating}. While \cite{solat2021distributed} introduces an HC method considering correlated uncertainties, it assumes predefined locations for renewable energy sources.

% optimization based
The optimization-based hosting capacity estimation have also been proposed in prior works. In this formulation, the objective is to maximize the installed capacity of DERs, subject to power flow equations and network operational constraints, such as voltage limits. This approach has been used in works \cite{home2022increasing,zhang2022increasing,cho2023stochastic}. However, one limitation of this method is that the optimization algorithm selects the location and size of DER in a way that maximizes the objective while satisfying all constraints. Hence, the resulting solution represents one possible optimal configuration and skips all other possible configurations that can potentially violate system constraints. This approach assumes that the system operator has control over DER placement and sizing, which might not reflect real-world scenarios where it is driven by the end-user decisions. As a result, the optimal solution obtained from the optimization problem might not reflect realistic DER placement scenarios, potentially leading to over-estimation of the true hosting capacity.

The HC can also be evaluated by methods other than load flow and optimization-based frameworks. An alternative approach to estimate  HC has been proposed in \cite{talkington2023measurement} using smart meter data and voltage sensitivities. However, the HC evaluated here is only for one bus at a time. Further, authors in \cite{munikoti2022novel,lliuyacc2024enhanced} have utilized voltage sensitivities to estimate the over-voltage violations and the HC for the network. However, the sensitivity methods include linearization of the power system model and may not accurately capture the non-linear behavior of the network under all conditions. Authors in \cite{li2023improved} have developed the HC problem as an optimization model that finds the locations that tend to be the worst for the DG integration and then maximizes the DG size for these locations. However, it leads to a conservative estimate of the HC. Authors in \cite{zhang2023model} have proposed a distributionally robust chance constraint (DRCC) optimization method to construct a feasible region for PV integration, given the set of candidate locations. In this work, the operational limits on voltage are converted to a DRCC model satisfying a given probability. Further, the PV output uncertainty is modeled by a data-driven Wasserstein-based ambiguity set. Authors in \cite{zhang2024analytical} have further utilized the DRCC-based feasible region construction for a net-zero distribution system having battery energy storage systems.

% Smart inverter
Apart from the estimation of the HC, there is a problem of increasing HC through additional measures, such as voltage control. This can include, for example, reactive power support from PV inverters. Numerous papers confirm a smart control of PV inverters that can help regulate nodal voltages, thus increasing the HC of a grid \cite{pareek2020optimal,jiao2024analytical}. Hence, in this work, we also use our proposed framework to analyze the impact of different voltage control approaches on the network's HC.

Considering the large amount of time taken by MCS-based frameworks, we aim to develop a novel HC estimation methodology using a chance-constrained optimization (CCO) approach based on the Gaussian Process regression (GPR) and logistic regression (LoR). There are multiple advantages of using CCO approach over robust optimization and scenario-based optimization. The robust optimization method ensures feasibility in all scenarios leading to a conservative solution and is computationally difficult \cite{erdougan2006ambiguous}.  On the other hand, CCO only requires satisfying system constraints with a certain confidence level, making it more practical \cite{geletu2013advances}. Scenario-based optimization struggles with the accuracy-complexity trade-off, where too many scenarios slow down the process, while too few lead to poor results.

GPR is a well-known machine learning algorithm known for its ability to effectively approximate smooth non-linear functions while providing a favourable balance between accuracy and complexity~\cite{liu2020gaussian, schulz2018tutorial}. The GPR-based frameworks have been used before to learn power flow~\cite{9552521}, optimal power flow~\cite{mitrovic2023data, pareek2020gaussian}, and battery health prediction~\cite{eleftheriadis2024comparative}.

The motivations for using CCO approach along with GPR and LoR are as follows:
\begin{enumerate}[label=\roman*.]
    \item 
\textbf{Risk management}: 
The chance-constrained framework will allow us for flexible risk management by controlling the probability of constraint violations. This adaptability is crucial for the distribution system operators, enabling them to quickly tailor the risk tolerance according to their specific needs.

\item \textbf{Uncertainty quantification}: 
 Instead of giving a single deterministic output, GPR provides a predictive mean and variance for each test point. The variance quantifies the uncertainty in the prediction.  
 % This is crucial for chance-constrained assessments, where calculating the probability of constraint violations or the risk index is essential. 
%which is useful for assessing the reliability of predictions. 
Similarly, LoR can also classify whether a given PV penetration level will lead to over-voltage or not with a certain probability.

\item 
\textbf{Cost-effectiveness}: The proposed models are computationally efficient and scalable with advanced techniques, such as sparse GPs and variational inference. By using these techniques, we can process and analyze large data, corresponding to a large number of possible scenarios, allowing for enhancing the accuracy and reliability of chance-constrained HC assessments.

\item 
\textbf{Flexibility and interpretability:} 
GPR and LoR are flexible and interpretable modeling techniques .
%that can be adapted to different types of data and systems. 
Especially, the GPR, being non-parametric, can accommodate any form of input uncertainty without requiring the knowledge of the uncertainty distribution. %Further, the use of GP-based and logit-based CCO methods will allow for flexibility and interpretability in the estimation as they can be expressed in terms of simple forms, see Section II for more details.

\item \textbf{Sensitivity Analysis:} Although HC estimation is a planning problem, the computational time is crucial for sensitivity analysis. For example, to assess HC reduction if bus 4 is unavailable for solar PV, Monte Carlo methods require a full re-run. In contrast, the proposed method uses probabilistic machine learning techniques like GPR and LoR, which require fewer samples and provide quicker results.

\end{enumerate}

By developing and validating a novel methodology for estimating the hosting capacity, this research will contribute to the development of a more sustainable and resilient energy future. The findings of this research will be valuable to stakeholders in the power sector, including policymakers, regulators, utilities, and renewable energy developers. 

\color{black}
The main contributions of the manuscript are as follows:
\begin{enumerate}
    \item Develop a novel chance-constrained HC estimation technique that accounts for both location and operational uncertainties. We also incorporate the correlation between the load demand and the PV generation using the copula theory.
    % These techniques can lead to a more realistic, accurate, and cost-effective estimation of the HC while also ensuring risk management or operational reliability awareness.   
    \item Develop a Gaussian process learning framework to quantify the impacts of the PV penetration level on the voltage distribution. 
    % Distribution grid operators can utilize these learning models to predict and estimate the probability of over-voltage violations, given any PV integration level.
    We also develop a logistic regression-based HC estimation model and compare its effectiveness with the GPR model. With the use of GPR and LoR, the chance-constrained HC estimation method can be reformulated in a very simple form which offers a fast and scalable solution.
    \item Analyze the effect of various voltage control techniques on the HC assessment such as droop-based PV inverter control and energy storage system control, using the proposed framework.
    % \item Conduct a comparison of different voltage control techniques for HC enhancement, such as reactive power, power factor, and energy storage system controls.   
\end{enumerate}

The rest of the paper is structured as follows. Section \ref{sec2} details the HC estimation problem. In section \ref{sec3}, we propose two main HC estimation methods with/without considering chance constraints of over-voltage issues, which are based on the GPR and LoR. Section \ref{sec4} presents the results and discussions. The data generation process along with PV inverter/ ESS control schemes are discussed in the Appendix.

\section{Problem formulation} \label{sec2}
We will start this section by introducing the probabilistic load flow (PLF) model to obtain samples of PV output and corresponding maximum voltage. We then formulate the chance-constrained optimization for the HC estimation. For improved readability, we have moved the details of the data generation process and the control methods to the Appendix which is present at the end of this paper.

\subsection{Load flow model}
The power flow equations can be defined as ~\cite{nguyen2018constructing}
\begin{align}
P_k + j Q_k & = V_k \sum_{l=1}^{N^b} Y_{kl} \,V_l\, \exp(-j \theta_{kl}) ,\quad k \in  \, \mathcal{L,G}. \label{eq:PF} 
\end{align}
The notations are defined below:
\begin{itemize}
    \item $P_k/Q_k$: active/reactive power injections at bus \textit{k}. 
\item $V_k, \theta_k$: voltage magnitude and phase at bus \textit{k}.
\item $\theta_{kl} = \theta_k - \theta_l$: phase difference between buses $k$ and $l$. 
\item $Y_{kl}$: element $(k,l)$ of admittance matrix Y. 
\item $ \mathcal{L}$ and $ \mathcal{G}$ are sets of all loads and generators in network.
\end{itemize}
Here, $N^b$ denotes the number of buses. The net power injection (net load) at bus $k$ can be given by: $ P_{L,k}:=P_{d,k}-P_{g,k}$. In case of additional components e.g. an ESS, the net injection can include power injection from that device too. 
We aim to capture the correlation between PV generation and demand. To achieve this, we use a copula function to model their joint cumulative distribution function (CDF). Later, the representative samples of load and demand are generated by drawing random samples from the estimated copula followed by inverse CDF calculation. For more details on the simulation setup and data generation, please refer to the Appendix. 

\subsection{Problem formulation of the HC estimation }
In this section, we first briefly introduce the HC estimation problem. Let $X$ denote the level of renewable penetration defined as the ratio of installed PV capacity to the system peak load demand power $\overline{P}_d$ (MW):
\begin{align} \label{eq:def_x}
    X:=\sum \limits_{k\in PV^{loc}}\mathcal{C}_k\Big/ \overline{P}_d, %\frac{\sum \limits_{k\in N_G}C_k}{\overline{P}_d},
\end{align}
where $\mathcal{C}_k$ is the PV rated capacity (MW) installed at bus $k$, and $PV^{loc}$ denotes the set of potential buses where PV units may be installed. 
The hosting capacity (HC) can be estimated by considering the following maximization problem \cite{chang2022equilibrium}:
\begin{subequations} \label{eq:det_hc}
\vspace{-2mm}
\begin{align} 
  \text{HC}:= & \max \limits_{X \in (0,1)} X  \label{eq:max_X} \\
    \text{s.t.}  & \qquad \textrm{Power Flow} \,\, \eqref{eq:PF}, \\
    %\,\,P_k + j Q_k & = V_k \sum_{l=1}^{N^b} Y_{kl} \,V_l\, \exp(-j \theta_{kl}) ,\quad k \in  \, \mathcal{L,G}, \label{eq:PF} \\
  & P_{g,k}^2 + Q_{g,k}^2 \leq S_{g,k}^2  ,\quad  k \in \, \mathcal{G},  \label{eq:cap_curve}\\
  & S_{g,k} \leq \mathcal{C}_k ,\quad k \in PV^{loc}, \\
  & \mathcal{C}_k \leq \overline{\mathcal{C}}_k ,\quad k \in PV^{loc},  \label{eq:ck}\\
   & \underline{V} \leq V_k \leq \overline{V},  \quad  k \in \, \mathcal{L},  \label{eq:no-chance}\\
%& \underline{\theta} \leq \theta_{kl} \leq \overline{\theta},  \quad  k,l \in \, \xi \\
&  \underline{s_{kl}} \leq s_{kl} \le \overline{s_{kl}},  \quad  k,l \in \, \mathcal{B}.  \label{eq:S}
\end{align}
\end{subequations} 
Here, $s_{kl}$ is the apparent power flow between bus $k$ and $l$ with $\mathcal{B}$ being the set of lines (branches) in the network. Constraint \eqref{eq:cap_curve} denotes the capability curve of the PV inverter with $S_{g,k}$ representing the apparent power through the PV unit at bus $k$. $\mathcal{C}_k$ is the rated capacity of the PV unit to be installed at bus $k$ with $\overline{\mathcal{C}}_k$ being the given upper limit. In the presence of other devices such OLTC, energy storage, etc. the respective device model can be added~\cite{ali2021maximizing, wu2022robust} \footnote{It should be noted that in the presence of other control devices such as capacitor banks, OLTC, etc., more constraints can be added. However, the objective of this work is to develop a generalized HC estimation framework instead of finding the optimal set-point for control devices. Hence, we have ignored them. However, they can be added to the proposed framework without any difficulty.}. Underline and overline represent the lower and upper operational limit on the variable, respectively. Later, in the proposed framework, we do not consider explicitly the under-voltage since the maximum PV penetration is limited by over-voltages due to reverse power flow~\cite{nguyen2015voltage}. However, this can be added to the problem formulation at almost no additional cost.  The counterpart on the line flow limits can be treated similarly.

\subsection{Formulation of CC-HC estimation}
In this section, we reformulate the deterministic HC problem (i.e. \eqref{eq:det_hc}) into the chance-constrained optimization model. Firstly, the stochastic nature of normalized power consumption $P_{dn}$ at each bus is modelled using normal distribution $P_{dn} \sim N(\mu_{d},\sigma_{d}^2)$. We further model the uncertain solar PV output using \textit{Beta} distribution $P_{gn} \sim \textit{Beta}(\alpha_{pv},\beta_{pv})$. The probability density function of PV generation is found to be well approximated by a beta distribution~\cite{ettoumi2002statistical}. 
Note that these normalized variables ($P_{dn}, P_{gn}$) can be multiplied with base values to obtain load $P_d$ and PV output $P_g$.

% Typically the HC of a network is limited by the over-voltage occurring due to the reverse power flow conditions.
We reformulate the hard limits on voltage (i.e. \eqref{eq:no-chance}) to chance-constrained formulations as~\cite{zhang2011chance}:
\begin{align}
    \mathbb{P}\big( V_{max}\le \overline{V} \big| X =x \big) \ge 1 - \beta,
\end{align}
where $V_{max}:=\max \limits_{1 \le k \le N^b } \{V_k\}$. It should be noted that $\overline{V}$ is the operational limit on bus voltages (i.e. 1.05 p.u.). $\mathbb{P}$ is the probability and $\beta$ is the risk level ($\beta \in (0,1))$. 

The chance-constrained HC estimation (CC-HC) can be formulated as follows:
\begin{subequations} \label{eq:cc_hc}
\begin{align} 
  &\text{CC-HC}:= \max \limits_{x \in (0,1)} x   \\
    \text{s.t.}\  & P_k^\omega + j Q_k^\omega  = V_k^\omega \sum_{l=1}^{N^b} Y_{kl} \,V_l^\omega\, \exp(-j \theta_{kl}^\omega) ,\  k \in \mathcal{L,G},   \\
  & (P_{g,k}^\omega)^2 + (Q_{g,k}^\omega)^2 \leq (S_{g,k}^\omega)^2,  \  k \in \, \mathcal{G},  \\
  & S_{g,k}^\omega \leq \mathcal{C}_k^\omega ,\quad k \in PV^{loc}, \\
  & \mathcal{C}_k^\omega \leq \overline{\mathcal{C}}_k ,\quad k \in PV^{loc},  \\
   & \mathbb{P}(V_{max} \leq \overline{V}| X =x) \geq 1 - \beta,
\end{align}
\end{subequations} 
where $\omega \in \Omega$ represents a scenario from the uncertain set.

Solving \eqref{eq:cc_hc} is difficult due to multiple reasons including computation intractability, uncertain location and size of future PV units, uncertain demand and PV output, etc. Furthermore, the non-linear nature of power flow equations adds an additional layer of complexity to the problem. Different models of power flow such as the DistFlow models can be used that reduce the complexity. However, the main challenge in solving this optimization problem is that the future locations of the solar PV units are uncertain, making the location a random variable.

\begin{figure}
    \centering
    \includegraphics[width= 0.98\linewidth]{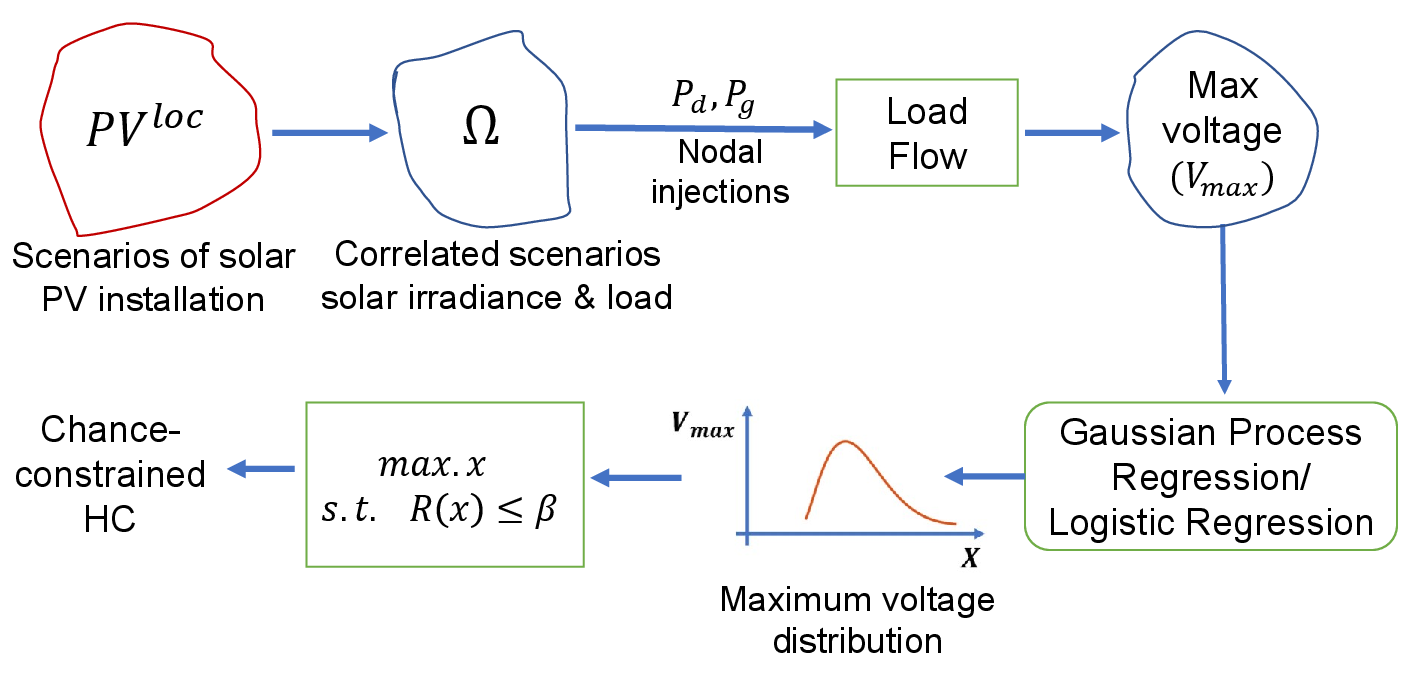}
    \caption{Process Flowchart for the proposed HC estimation framework.}
    \vspace{-0.5 cm}
    \label{fig:sys_arc}
\end{figure}

Instead of directly solving the stochastic optimization problem \eqref{eq:cc_hc}, we adopt a different approach with three stages. \textbf{In stage 1}: We generate random scenarios for solar PV installation along with the correlated scenarios for solar PV output and load. Then we run the probabilistic load flow \eqref{eq:PF} to obtain a training dataset consisting of the PV penetration ($X$) as input and the maximum of all bus voltages ($V_{max}$) as output (see Appendix for more details on how we generate the data). \textbf{In stage 2}: using this dataset, we construct a predictive function $V_{max}=f(X)$ (through Gaussian Process modeling).
Finally, \textbf{in stage 3}: we define a risk index based on voltage distribution as:
\begin{align}
    R(x):=\mathbb{P}(V_{max}> 1.05 \big| X=x) \label{eq:prob_vmax},
\end{align}
Once the model of \eqref{eq:prob_vmax} is known, the HC problem i.e.  \eqref{eq:cc_hc} is transferred to:
\begin{align}
  \text{CC-HC}:= & \max \limits_{x \in (0,1)} x \quad   \text{ s.t. } \quad R(x)  \le \beta,    \label{eq:chance-risk}
\end{align}
for some risk level $\beta \in (0,1)$. In this study, we consider several risk levels, including $\beta = 1\%, 5\%,$ and $10\%$, and we set the voltage limit $\overline{V} = 1.05$ (p.u.). The overall process flow is shown in Fig. \ref{fig:sys_arc}. 

However, the computation of the conditional probability \eqref{eq:prob_vmax} is a nontrivial task as we do not know the exact form and it is impossible to consider the whole possible outcomes of the continuous random variables $X$ and $V_{max}$. Thus, it motivates us to leverage probabilistic machine learning models with available data. Specifically, we propose to use GPR and LoR models in this work. 

% \color{red}  The constraint \eqref{eq:chance-reliability} serves as an \textit{operational reliability index}, allowing us to determine the maximum level of PV penetration that ensures system safety with a high-reliability level. \footnote{While maintaining appropriate voltage levels is crucial for operational reliability, it is also essential to consider other factors such as thermal limits, frequency stability, and protection coordination to ensure a comprehensive assessment of grid reliability. However, this work is restricted to present results with violated voltage concerns only, other factors can be included in the same manner. } \color{black}

\vspace{-1em}
\section{Proposed methods} \label{sec3}
In this section, we first develop the GP-based chance-constrained HC estimation method. Recall that the Gaussian process is a collection of random variables, any finite number of which have a joint multivariate Gaussian distribution. While GP is a general framework for modelling distributions over functions, GPR is a specific application of GPs for regression tasks \cite{williams2006gaussian}. Unlike traditional regression methods, GPR does not assume a specific functional form, but uses a GP as a prior over functions and combines it with new input data to make predictions, along with uncertainty quantification ~\cite{mitrovic2023data}. Thus, GPR is a non-parametric probabilistic model and capable of handling complex non-linear input-output relationships. Later, we also develop the logistic regression-based CC-HC estimation method and compare the efficiency of both methods in section \ref{sec4}.

\subsection{GPR-based HC estimation methods}
Suppose we have a training sample of $(X,V_{max})$ denoted by $(x_1,\ldots,x_n)^T$ and $(v_{max,1},\ldots, v_{max,n} )^T$, where $n$ is the sample size. Recall that $X$ here refers to the PV penetration level (refer to \eqref{eq:def_x}) and $V_{max}$ refers to the maximum voltage among all buses. Without loss of generality, let us center the output $V_{max}$ to have a zero mean by subtracting it from its sample mean ${m}:=\frac{1}{n}\sum\limits_{i=1}^n v_{max, i}$ as $y_i:=v_{max,i} - m$. We are seeking a predictive function $f(x)$ such that:
\begin{align}
    y_i=f(x_i) +\varepsilon_i, \label{eq:model}
\end{align}
where $\varepsilon$ denotes the noise term with zero mean and constant variance $\sigma_\varepsilon^2$. Here, we model $f(x)$ as a GP defined as a collection of latent variables indexed by the input $x$ and having joint multivariate Gaussian distribution \cite{williams2006gaussian}. In GPR, it combines prior and likelihood probabilities to derive posterior and predictive distributions, see e.g. \cite{williams2006gaussian} in detail. Suppose the new prediction is $y^*$ using a new input $x^*$, conditional on the training observation $(x_i,y_i), \, i=1,\ldots,n$. This can be done by using the key predictive distribution of $f^*=f(x^*)$ determined by:
 
\begin{align}
    %& R|(R_1=r_1,\ldots,R_n=r_n, X_1=x_1,\ldots,X_n=x_n, X=\widetilde{x})  \\
  p( f^*\big| \bm{x},\bm{y},x^*)  & \simeq \mathcal{N}(\mu_f(x^*),\sigma_f^2(x^*)), 
\end{align}
with the mean and variance functions defined by \cite{williams2006gaussian}:
\begin{align}
\mu_{f}(x^*)&:=\bm{K}_*^T\big(\bm{K}+\sigma_\varepsilon^2\bm{I}\big)^{-1}\bm{y}  \label{eq:mean},\\
    \sigma_{f}^2(x^*) &:= \bm{K}_{**}-\bm{K}_*^T\big(\bm{K}+\sigma_\varepsilon^2\bm{I}\big)^{-1}\bm{K}_* \label{eq:var}.
\end{align}
where $\bm{I}$ is the unit matrix of size $n \times n$, $\bm{K}$ denotes the training kernel matrix of size $n \times n$, $\bm{K}_*$ the training-testing kernel matrix of size $n \times 1$, and $\bm{K}_{**}$ is the testing kernel matrix of size $1 \times 1$, which can be explicitly calculated as follows, for example, if using "\textit{squared exponential kernel}": 
\begin{align}
    \bm{K}(x_i,x_j) &= \lambda^2 \exp\big(-\frac{(x_i-x_j)^2}{\tau^2}\big), \quad i,j=1,\ldots, n,\nonumber \\
    \bm{K}_*(x_i,x^*) & = \lambda^2 \exp\big(-\frac{(x_i-x^*)^2}{\tau^2}\big), \quad i=1,\ldots, n, \nonumber  \\
    \bm{K}_{**}(x^*,x^*) &= \lambda^2 \exp(-\frac{(x^*-x^*)^2}{\tau^2}\big)=\lambda^2, \nonumber
\end{align}
where the variance of the noise $\sigma_\varepsilon^2$, the hyper-parameters $(\lambda^2,\tau^2)$ of the kernel function are estimated from the training data $(x_i, y_i)_{i=1}^n$ by maximizing the log marginal likelihood,  see (2.30) in \cite{williams2006gaussian}.  

To get back the original scale prediction for the new output $v^*_{max}$  given on the new input $x^*$, we simply calculate the mean $\mu(x^*)$ and variance $\sigma^2(x^*)$ functions of $v^*_{max}= y^*+m$, with $ y^*=f(x^*)+\epsilon$:
\begin{align}
    \mu(x^*)= \mu_f(x^*)+m, \quad  \sigma^2(x^*)= \sigma^2_f(x^*) +\sigma_\varepsilon^2. \label{eq:mu_sigma}
\end{align}

% In practice, $\mu(x^*)$ is used as a point prediction, while the $(1-\alpha)\times 100$ (\%) prediction interval (PI) of $V_{max}$ can be constructed by:
In practice, $\mu(x^*)$ is used as a point prediction. Suppose we desire a prediction interval with a confident level of $(1-\alpha)\times 100$ (\%). This can be evaluated as
\begin{align}
    \mu(x^*) \pm z_{\alpha/2} \sigma(x^*), \label{eq:PIs}
\end{align}
where $z_{\alpha/2}$ is the upper quantile of standard normal distribution at level $\alpha/2$. 

\subsubsection{GPR-based HC estimation method for mean and confidence interval bounds}

By utilizing the learned GPR, we now can reformulate the HC estimation considering the over-voltage issue in \eqref{eq:no-chance} as follows:
    \begin{align}
    \text{GP-WoCC-HC}:=& \max \limits_{x^* \in (0,1)} \{x^*\}  \quad \text{s.t.}  \quad \mu(x^*) \le 1.05,   \label{eq:GP-WoCC-Vmax}
\end{align}
where $\mu(x^*)$ is defined as in \eqref{eq:mu_sigma}. In this work, we shall refer to the solution of \eqref{eq:GP-WoCC-Vmax} as the mean \textbf{GP-WoCC-HC}, i.e., the average HC estimation resulting from the GP learning without considering chance constraints. 

Further, the GPR approach can also offer us both the lower and upper bounds for the  HC estimation by utilizing the prediction interval (PI) \eqref{eq:PIs}:
\begin{align}
   \text{HC bounds}:= & \max \limits_{x^* \in (0,1)} \{x^*\}  \nonumber \\
   &\text{s.t. } \, \mu(x^*) \pm z_{\alpha/2} \sigma(x^*) \le 1.05,   \label{eq:GP-WoCC-Vmax-bounds}
\end{align}
for given a confident level: $(1- \alpha)\times 100\%$.

\subsubsection{GPR-based CC-HC estimation method}
Now, to develop the HC estimation with chance constraints, we calculate the probability of constraint violation or the risk index in \eqref{eq:prob_vmax} as:
%develop the upper and lower bounds by using the prediction intervals (PI) i.e. \eqref{eq:PIs} as
% Further, by applying the learned GPR for computing the chance-constrained voltage as \eqref{eq:prob_vmax}:
\begin{align}
    \mathbb{P}(V_{max}> 1.05 \big | X=x^*) =1-\Phi\Big(\frac{1.05-\mu(x^*)}{\sigma(x^*)}\Big). \label{eq:risk_index}
\end{align}
Then, we can reformulate the chance-constrained HC estimation in \eqref{eq:chance-risk} based on this GP learning (\textbf{GP-CC-HC} for short) as:
\begin{align}
   \text{GP-CC-HC}:= & \max \limits_{x^* \in (0,1)} \{x^*\}  \nonumber \\
   &  \text{s.t. }  \,  \mu(x^*)+\sigma(x^*)\Phi^{-1}\big(1-\beta \big) \le 1.05,    \label{eq:GP-CC-Vmax}
\end{align}
for some risk level $\beta  \in (0,1),$ where $\mu(x^*)$ and  $\sigma(x^*)$ are the predicted mean and standard deviation of the  $V_{max}$ through the learned GPR defined as in \eqref{eq:mu_sigma}, and $\Phi^{-1}$ denotes the inverse CDF (quantile function) of the standard normal distribution. 

\subsubsection{A relationship between GP-WoCC-HC and GP-CC-HC estimation methods}
Note that if we take $\beta=\alpha/2$, then we have
\begin{align}
    \Phi^{-1}\big(1-\beta \big) = \Phi^{-1}\big(1-\alpha/2 \big) =z_{\alpha/2},
\end{align}
and thus, the \textbf{GP-CC-HC} problem in \eqref{eq:GP-CC-Vmax} with a risk level of $\beta$ is equivalent to the lower bound of the \textbf{GP-WoCC-HC} problem in \eqref{eq:GP-WoCC-Vmax-bounds} with the confidence level of $$(1-\alpha)\times 100 (\%)=(1-2\beta) \times 100 (\%).$$

Next, for a simple comparison purpose with the proposed GP learning, we provide another approach using the so-called logit learning.

\subsection{LoR-based CC-HC estimation method}
Another approach to compute the risk \eqref{eq:prob_vmax} is via logistic regression. To do so, we begin by defining the following violation variable:
\begin{align}
    \widetilde{V} = \left\{ \begin{array}{cc}
         1, &  \quad \mbox{if} \quad V_{max}> 1.05, \\
        0, &   \quad \mbox{if} \quad \mbox{otherwise.}
    \end{array} \right.
\end{align} 
Then we can learn the conditional violation probability by using the logistic regression model of the form:
\begin{align}
   \mathbb{P}(\widetilde{V}=1 | X=x^*)= 1/\big(1+\exp(-\widehat{b}_0-\widehat{b}_1 x^*)\big), \label{eq:logit-prob} % \frac{1}{1+\exp(-b_0-b_1x)}, \label{eq:logit-prob}
\end{align}
where $ \widehat{b}_0$ and $\widehat{b}_1$ are the estimated parameters of the logistic regression using the training data $(x_i,\widetilde{v}_i),\,i=1,\ldots,n$. We now can determine the chance-constrained HC such that $\widetilde{R}(x^*):=\mathbb{P}(\widetilde{V}=1  \big | X=x^*) \le \beta $, i.e.,
\begin{align}
    & \max \limits_{x^* \in (0,1)} \{x^*\} \,\text{ s.t. }  1/\big(1+\exp(-\widehat{b}_0-\widehat{b}_1 x^*)\big) \le \beta, \label{eq:logit-CC-Vmax} %\frac{1}{1+\exp(-b_0-b_1x)} \le \beta,    
\end{align}
for some risk level $\beta  \in (0,1)$. Or equivalently, we have in terms of the logit representation: 
\begin{align}
   \text{Logit-CC-HC}:= & \max \limits_{x^* \in (0,1)} \{x^*\} \nonumber \\
    & \quad \text{ s.t.} \quad \widehat{b}_0+\widehat{b}_1 x^* \le \text{logit}(\beta),  \label{eq:logit-CC-Vmax2}
\end{align}
where $\text{logit}(\beta):=\log\Big(\frac{\beta}{1-\beta}\Big)$, for some risk level $\beta  \in (0,1)$.  We shall refer to this approach as the \textbf{Logit-CC-HC} method, i.e. logit-based chance-constrained HC estimation.

\section{Results and discussions} \label{sec4}
In this section, we evaluate the proposed models. %After the trained GPR and LoR models have been confirmed with good performance, we then apply them to solve the proposed chance-constraint optimization for determining the HC. 
The load flow analysis was performed using the Matpower toolbox . The data generation process along with the PV inverter/ ESS control schemes are given in the Appendix.

\subsection{Test system}
A modified IEEE 33-bus \cite{baran1989network} and 123-bus \cite{bobo2021second} networks were used as test systems. The 33-bus network contained 2 capacitor banks of size 0.4 MVar at buses 18 and 33. The peak load in this network is 3.715 MW, 2.3 MVar with a base voltage of 12.66 kV. The 123-bus network was equipped with capacitor banks of size 0.4, 0.4, and 0.2 MVar at bus 51, 61, and 64, respectively.  
% The network also contains a voltage regulator between buses 67 and 160 with a tap ratio 1.03.
The peak load in this network is 3.51 MW, 1.93 MVar with a base voltage of 4.16 kV. In both the networks, the point of common coupling (PCC) voltage was set at 1.03 p.u. to ensure that the lowest bus voltage is above $0.95$ p.u. under the peak load condition. Simulations were performed on a computer with an Intel Xeon processor 3.7 Ghz, 16 GB RAM, and using MATLAB 2019a. 

We considered 3000 location-size scenarios of PV installation. Further, we considered 4 representative load-generation profiles for each location-size scenario resulting in a total of 12000 scenarios to be analyzed using load flow analysis. We selected the period between 12 to 1 p.m. for analysis as it usually observes high sunshine and low residential demand, thus capturing the high bus voltages caused by solar PV~\cite{ukil}. This also ensures that the resulting HC estimate will not cause over-voltage violations for any load pattern on weekdays or weekends.

The historical PV output data was collected from the Photovoltaic Geographical Information System (PVGIS) toolbox \cite{huld2012new} for the year 2020. The data was fitted on \textit{Beta} distribution and the obtained shape parameters ($\alpha_{pv}$ and $\beta_{pv}$) were approximately equal to 15 and 6. It should be noted that $\beta_{pv}$ is restricted to PV output and should not be confused with risk parameter $\beta$. Further, the normalized load was considered to be following a normal distribution with mean ($\mu_d$) and standard deviation ($\sigma_d$) equal to 0.5 and 0.025 p.u. respectively. The Pearson's correlation coefficient was considered as 0.15. The resulting 4 representative load-generation pairs obtained from Module 1 (see Appendix) are $[$(0.54-0.96), (0.52-0.95), (0.51-0.93), (0.47-0.92)$]$ with all values in per unit. The load and PV output at each bus were scaled using these values. The load flow analysis was performed for each scenario with and without considering the control action. 

\color{Black}
%\vspace{-2mm}
\subsection{Evaluating the proposed models} 

\begin{figure}[tb]
    \centering
    \includegraphics[scale=0.5]{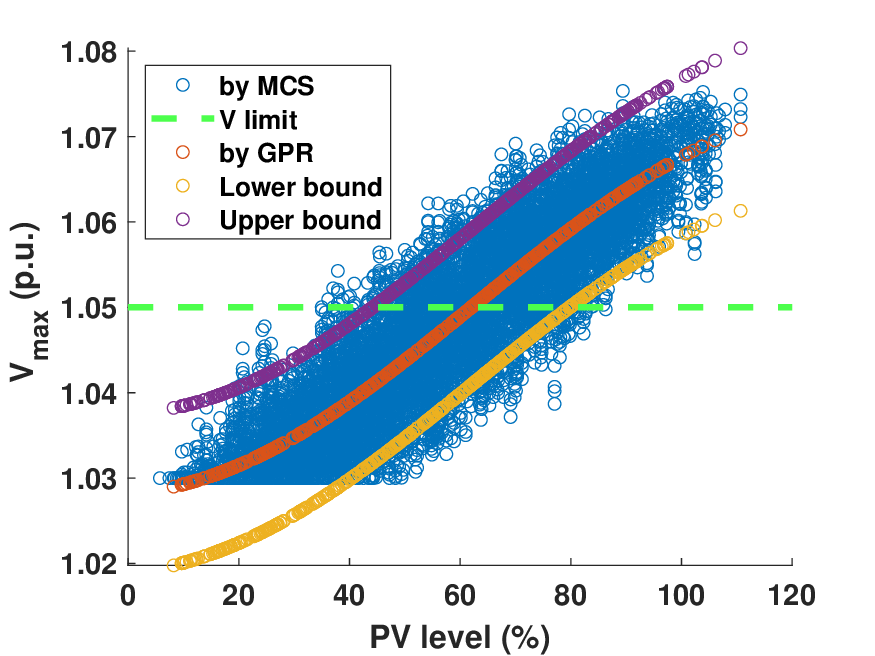}
    \vspace{-1mm}
    \caption{GPR used for predicting $V_{max}$ on IEEE 33-bus system, given different PV penetration levels, in comparison to Monte Carlo simulation (MCS). } 
    \vspace{-5mm}
    \label{fig:GPR_Vmax_33bus}
\end{figure}

Firstly, Fig. \ref{fig:GPR_Vmax_33bus} plots the trained GPR with 500 random samples of the PV level input, in comparison to the Monte Carlo (MC) simulation of 12000 samples on the IEEE 33-bus network. The rest of the 11500 samples will be used as testing data. As can be seen, the GPR can be used for point predictions as well as for providing the lower and upper bounds (95\% PI) of the maximum voltage output $V_{max}$. The error point predictions are small, with the mean absolute error (MAE) of only 0.0037 and  0.0021 (p.u.), performed on the IEEE 33-bus and 123-bus systems, respectively, as shown in Tab. \ref{tab:errors_analysis}. We also achieve high coefficients of determination $\bm{R}^2$, around $85.76\%-92.91\%$ regarding the regression task. Not only can GPR be applied for forecasting the continuous random variable $V_{max}$, but it can also be utilized for classifying the event experiencing the over-voltage. This can be done by calculating the risk index $R(x^*)$ as in \eqref{eq:risk_index} for GPR and \eqref{eq:logit-prob} for LoR. Specifically, given an integration of PV level $X=x^*$, we predict the likelihood of the over-voltage $R(x^*)$ and then classify it as an over-voltage violation if $R(x^*) > 0.5$. Tab. \ref{tab:errors_analysis} shows that the classification accuracy (\textbf{Acc.}) by the GPR model is around 89.96\% for IEEE 33-bus and 93.01\% for IEEE 123-bus, performed on the testing data of 11500 samples. The logistic regression models perform similarly on the classification task with an accuracy of 89.95\% % and 93.01\% %, respectively.  

\begin{table}[h]
    %\vspace{-2mm}
    \centering
    \caption{ Performance evaluations of GPR and LoR models on the testing data of 11500 samples. }
    \label{tab:errors_analysis}
    \begin{tabular}{c|c|cccc}
  \textbf{IEEE} &  \textbf{Model} &   \textbf{MAE} & \textbf{RMSE} & $\bm{R}^2$ & \textbf{Acc.} \\ \hline
   \multirow{2}{*}{33-bus}   & GPR  & 0.0037  & 0.0047 & 85.76\% & 89.96\% \\
       & LoR & -- & --  &  -- & 89.95\%\\ \hline
      \multirow{2}{*}{123-bus}   & GPR  &  0.0021 & 0.0025 & 92.91\% & 93.01\% \\
       & LoR & -- & --  &  -- & 93.01\%\\ \hline
    \end{tabular}
   \vspace{-4mm}
\end{table}

\begin{figure}[tb]
\centering
\begin{subfigure}{.75\linewidth}
  \includegraphics[width=\linewidth]{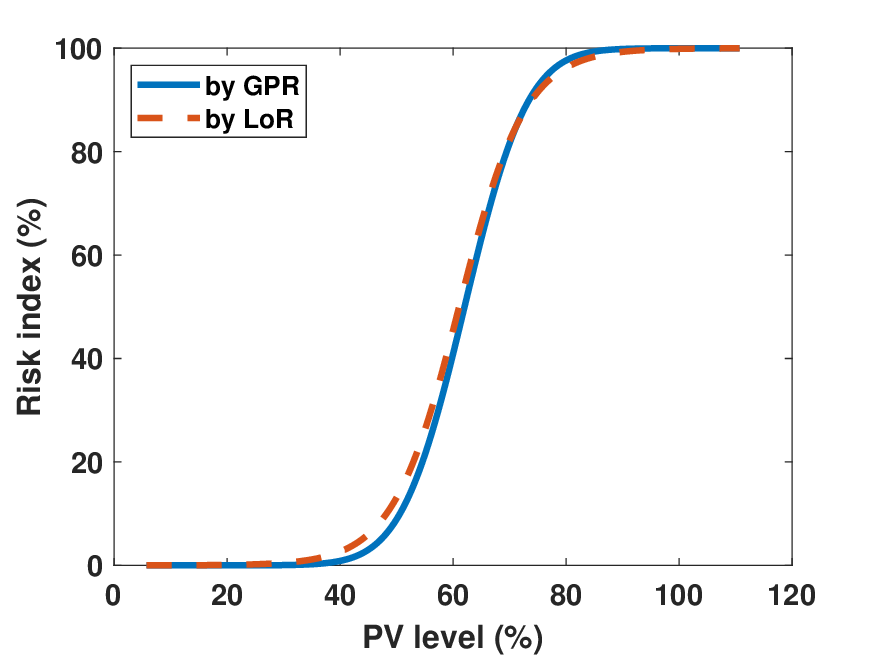}
  \caption{33-bus}
  \label{fig:prob_Vmax_33bus}
\end{subfigure}\hfill  
\begin{subfigure}{.75\linewidth}
  \includegraphics[width=\linewidth]{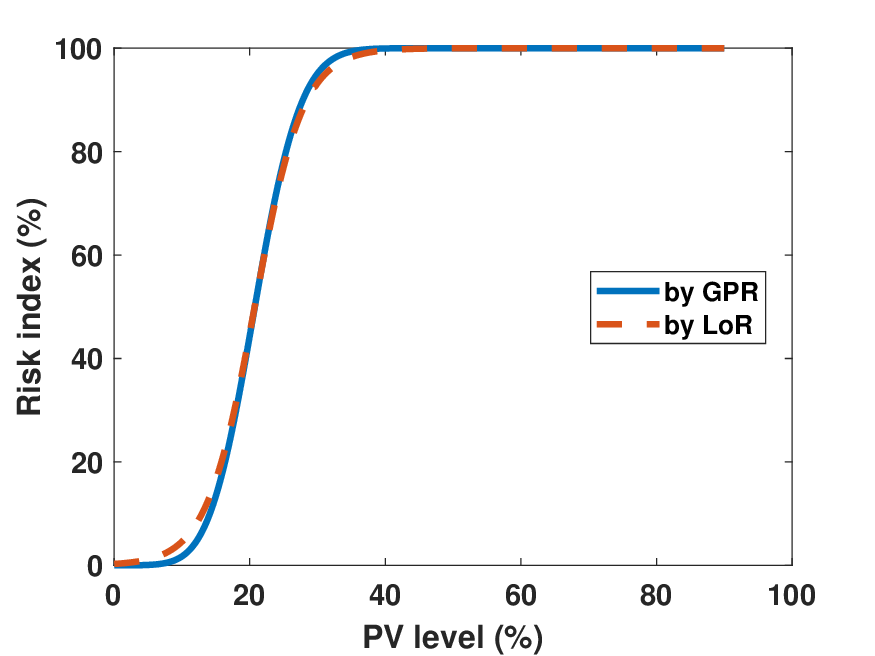}
  \caption{123-bus}
  \label{fig:prob_Vmax_123bus}
\end{subfigure}
 \caption{Comparison between GPR and LoR used for predicting the risk index $R(x)$ on IEEE 33- and 123-bus systems, given different PV penetration levels \textbf{(No control).}}
\vspace{-6mm}
\label{fig:prob_Vmax_33_123bus}
\end{figure}

Fig. \ref{fig:prob_Vmax_33_123bus} compares GPR and LoR in estimating risk curves for IEEE 33-bus and 123-bus systems. Both models show similar risk estimations, confirming the consistency of our proposed models in classifying over-voltage issues. However, it is important to note that the LoR model cannot predict maximum voltage output like the GPR model. Therefore, the LoR model is solely used for calculating chance-constrained PV hosting capacity, while GPR can be applied for estimating the mean and bounds of HC as well as chance-constrained HC, as discussed in the next section.

% We acknowledge that the performance of GPR and LoR can be influenced by the underlying probabilistic assumptions of the data. 
GPR is non-parametric learning method, so the performance does not depend strongly on the input distribution. GPR has proven efficient on real-world datasets \cite{cai2020gaussian,zeng2020prediction}, while LoR has been applied to actual water pipeline failure data \cite{robles2020prediction}. In case the samples are not drawn from the assumptions of the probabilistic models, then we can apply transformation techniques like isoprobabilistic transforms, or Nataf transform before training the models \cite{lebrun2009generalization}.  

\color{Black}
%\vspace{-3mm}
\subsection{Enhanced hosting capacity estimations with control techniques}

In the previous section, the effectiveness of the proposed GPR and LoR methods have been verified, as shown in Tab. \ref{tab:errors_analysis}. In this section, we apply the GPR to estimate the mean and confident intervals for the HC, using the GP-WoCC-HC method proposed in \eqref{eq:GP-WoCC-Vmax}-\eqref{eq:GP-WoCC-Vmax-bounds}, which does not consider the chance constraint. Subsequently, we present the results of chance-constrained HC using both GPR and LoR approaches, as proposed in \eqref{eq:GP-CC-Vmax} and \eqref{eq:logit-CC-Vmax2}, respectively. 

We also estimated HC considering multiple control methods. The first two methods control the PV inverter's output through Volt-Var (Q) control and power factor (PF) control. The droop characteristics and the control methodology are described in Module 4 of the Appendix. 
% For the 33-bus network, we evaluated HC under droop-based Q control and PF control. 
For the 123-bus system, we additionally evaluated HC under energy storage system (ESS) control. 
% The third method is also a droop-based method which charge-discharges the energy storage system (ESS) according to the nodal voltages. 
For this case study, the 123-bus was considered to be equipped with six ESS with power ratings of 150 (kW) each and located at bus $[$50, 70, 72, 80, 105, 110$]$. We assume that the control actions comply with the ESS state-of-charge limits.
 
Tab. \ref{tab:HC-GP-bounds} presents the estimated mean values of the PV HC for the IEEE 33-bus and 123-bus networks, along with the lower and upper bounds of HC at a 95\% confidence level. Additionally, the effectiveness of enhanced control techniques, including PV inverter control and ESS control, are shown to increase the HC compared to results without control actions. Within the settings in this work, our findings indicate that PF control is more efficient than Q control in enhancing the HC in the IEEE 33-bus network, while the ESS technique outperforms PV inverter control in the IEEE 123-bus system. For instance, in the IEEE 33-bus network, the HC can be increased from a mean of 62.02\% of the peak load of 3.715 (MW) to 71.12\% on average with Q control, while PF control can improve the HC up to 74.11\%. However, we observed that the PF control is only slightly better than the Q control in maximizing the HC mean value in the IEEE 123-bus network, but it also showed a great variance with the bounds ranging from a minimum of 18.89\% to a maximum of 85.12\%. The reason could be that the IEEE 123-bus is more compact with shorter branches than the IEEE 33-bus network. 

\begin{table}[!ht]
    \centering
       %\vspace{-2mm}
    \caption{GP-WoCC-HC method for estimating means and $95\%$ confident interval bounds of PV hosting capacity levels (in terms of $\%$ of the peak load).}
    %\vspace{-2mm}
    \label{tab:HC-GP-bounds}
    \begin{tabular}{c|c|ccc}
   \textbf{IEEE} & \textbf{Control} & \textbf{mean} &   \textbf{lower} & \textbf{upper}  \\ \hline
     \multirow{ 3}{*}{33-bus} &  no  & 62.02\%  & 44.20\% & 79.79\% \\
        & Q & 71.12\% & 53.69\% & 92.38\% \\
        & \textbf{PF} & 74.11\% & 55.95\% &   98.71\% \\ \hline
        \multirow{ 3}{*}{123-bus} &  no  & 20.76\%  & 10.76\% & 31.81\% \\
        & Q & 36.22\% & 22.22\% & 54.85\% \\
        & PF & 37.31\% & 18.89\% & 85.12\% \\
        & \textbf{ESS} & 45.15\% & 31.30\% &  57.54\% \\ \hline
    \end{tabular}
    \vspace{-4mm}
\end{table}

Next, we present the estimation of chance-constrained HC, ensuring that the likelihood of over-voltage is at most risk levels of $\beta=1\%, 5\%$, and $10\%$. The results for all cases are shown in Table \ref{tab:HC-GP-logit}. As expected, the resulting CC-HC values are lower than the ones without the chance constraint due to the risk-averse nature of the optimization. For instance, at an extremely risk-averse level of $\beta=1\%$, the GPR-based CC-HC in the IEEE 33-bus is only 40.56\% of the peak load of 3.715 (MW). Meanwhile, the PF control technique can improve the GPR-based CC-HC up to 50.50\%. Furthermore, we find that the logit-based model mostly provides smaller estimations compared to the GPR approach. Nevertheless, as the acceptable risk levels increase, the results from both models become closer, as seen in the case of $\beta=10\%$ with the ESS control in the IEEE 123-bus, where the CC-HC is estimated to be 36.70\% and 37.24\% by the GP and logit learning approaches, respectively. One possible reason could be that as the risk level increases, the chance constraints become less restrictive. Note that our proposed methods take a total time of less than 3.24 seconds to generate 500 samples of the load flow, train the GPR and LoR models, and solve the proposed GP-CC-HC and logit-CC-HC optimizations for the estimations shown in Table \ref{tab:HC-GP-logit}.

\begin{table}[!ht]
    \centering
       %\vspace{-4mm}
    \caption{Comparison of GP-CC-HC and Logit-CC-HC methods for the estimation of chance-constrained PV hosting capacity levels.}
    %\vspace{-2mm}
    \label{tab:HC-GP-logit}
    \begin{tabular}{c|c|c|p{1.2cm}p{1.2cm}}
   \textbf{IEEE} & \textbf{Control} & \textbf{Risk level} $\beta$ &   \textbf{GP-CC-HC} & \textbf{Logit-CC-HC}  \\ \hline
     \multirow{9}{*}{33-bus} & \multirow{ 3}{*}{no} & 1\%  &   40.56\% & 34.19\% \\
       & & 5\% & 47.21\% & 43.84\% \\
       & & 10\% &  50.59\% &   48.21\% \\ \cline{2-5}
       & \multirow{ 3}{*}{Q} & 1\%  &   50.50\% & 45.72\% \\
       & & 5\% &56.43\% & 54.32\% \\
       & & 10\% &  59.59\% &   58.22\% \\ \cline{2-5}
       & \multirow{ 3}{*}{PF} & 1\%  &   52.68\% & 46.71\% \\
       & & 5\% & 58.76\% & 56.05\% \\
       & & 10\% &  62.02\% &   60.28\% \\ \hline
        \multirow{9}{*}{123-bus} & \multirow{ 3}{*}{no} & 1\%  & 8.96\% & 4.49\% \\
       & & 5\% & 12.31\% & 10.33\% \\
       & & 10\% & 14.13\% &  12.97\% \\ \cline{2-5}
        & \multirow{ 3}{*}{Q} & 1\%  &   20.22\% & 20.68\% \\
       & & 5\% & 24.06\% & 26.26\% \\
       & & 10\% & 26.35\% &  28.79\% \\ \cline{2-5}
       & \multirow{ 3}{*}{PF} & 1\%  &   16.57\% & 4.75\% \\
       & & 5\% & 21.01\% & 16.80\% \\
       & & 10\% & 23.66\% &   22.25\% \\ \cline{2-5}
       & \multirow{3}{*}{ESS} & 1\%  &   27.80\% & 28.66\% \\
       & & 5\% & 33.79\% & 34.57\% \\
       & & 10\% &  36.70\% & 37.24\% \\ \hline
    \end{tabular}
    \vspace{-4mm}
\end{table}

\begin{figure}[!ht]
    \centering
    \includegraphics[scale=0.46]{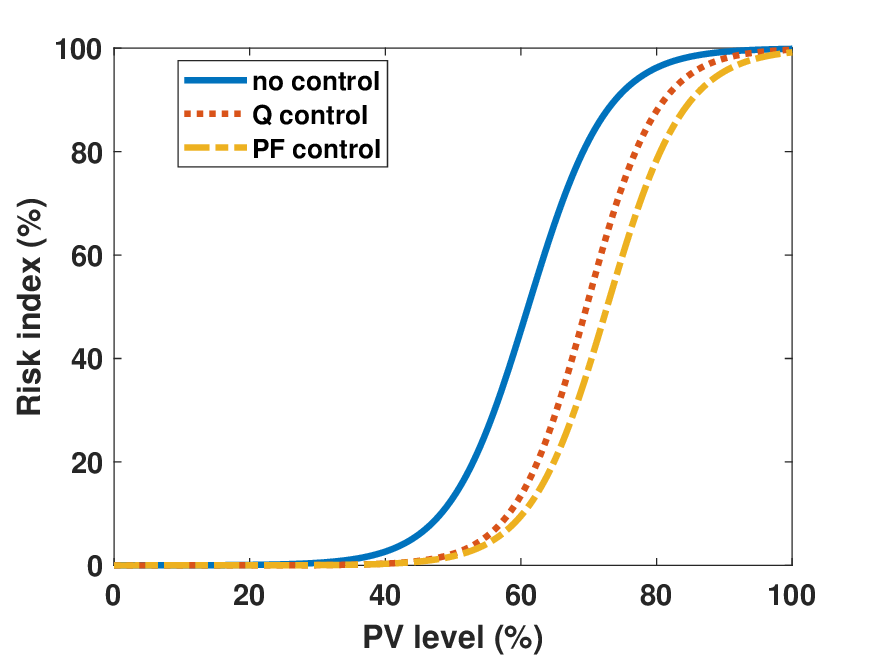}
    \vspace{-2mm}
    \caption{Risk curves by LoR for IEEE 33-bus system, considering the reactive power (Q) and power factor (PF) controls.} 
    \vspace{-5mm}
    \label{fig:logit_Vmax_risks_33bus}
\end{figure}

\begin{figure}[!ht]
    \centering
    \includegraphics[scale=0.46]{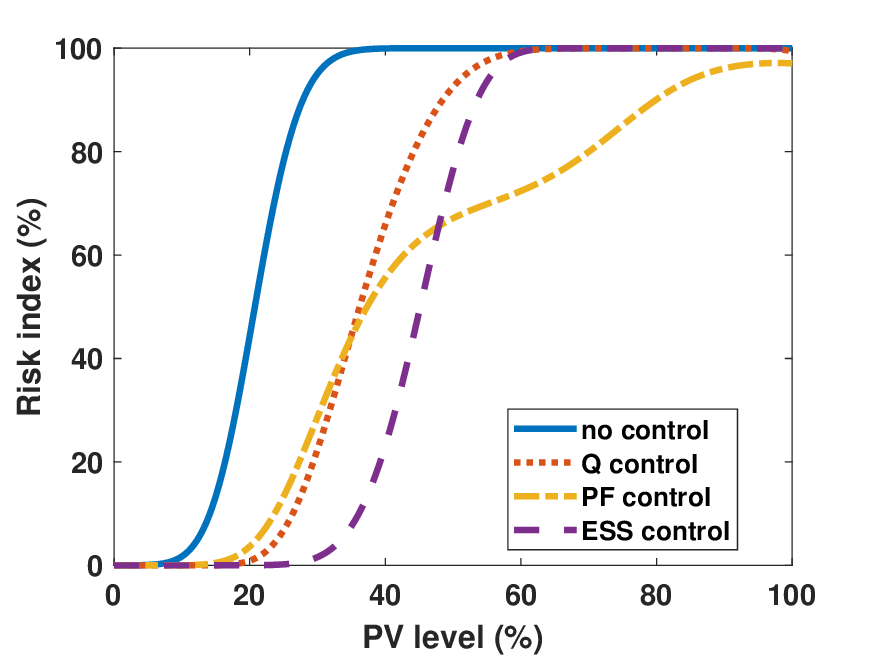}
     \vspace{-2mm}
    \caption{Risk curves by GPR for IEEE 123-bus system, considering the reactive power (Q),  power factor (PF), and energy storage system (ESS) controls.} 
    \vspace{-4mm}
   % \vspace{-3mm}
\label{fig:GP_Vmax_risks_123bus}
\end{figure}

Figures \ref{fig:logit_Vmax_risks_33bus} and \ref{fig:GP_Vmax_risks_123bus} display risk curves estimated by the LoR models for the IEEE 33-bus and by the GPR models for the IEEE 123-bus, respectively. These curves enable system operators to quickly determine the over-voltage risk at any PV penetration level in the networks, which is valuable for decision-making. Additionally, the figures demonstrate the effectiveness of Q, PF, and ESS control techniques in improving HC, as evidenced by the fact that at the same PV penetration level, the risks are reduced. 
% These risk curves and control techniques provide valuable insights for RES integration into distribution networks.
%\vspace{-0.5 cm}

\begin{figure}[!ht]
    \centering
    \includegraphics[scale=0.46]{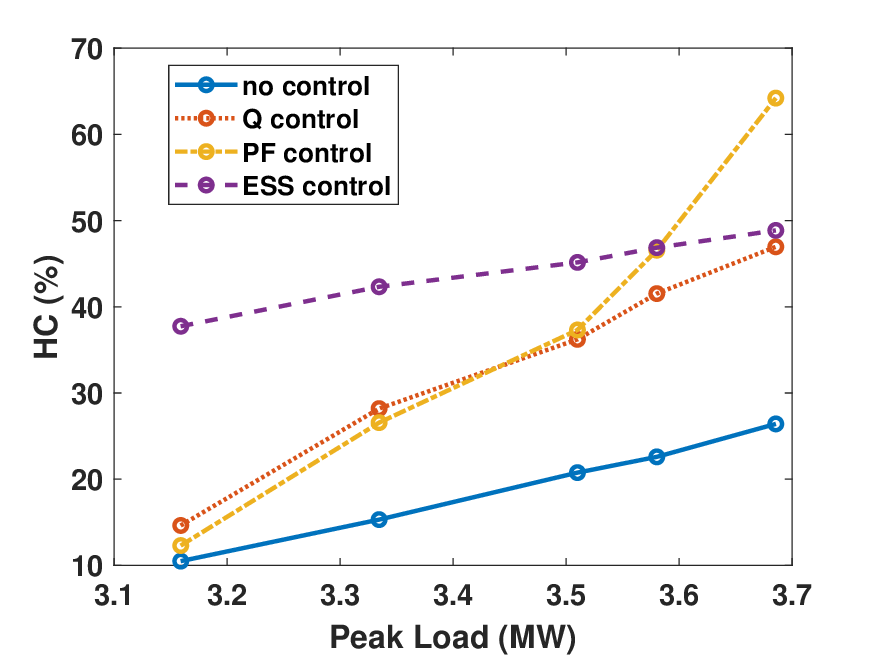}
     \vspace{-0.2 cm}
    \caption{Estimating the mean values of the HC for the IEEE 123-bus system, considering different peak loads.} 
    \label{fig:mean_HC_peakloads_123bus}
    \vspace{-6mm}
\end{figure}

We further examine the estimated HC under various peak loads, as depicted in Fig. \ref{fig:mean_HC_peakloads_123bus}. Our findings indicate that as peak loads in networks rise, so do HC values. This trend holds even when considering enhancements like reactive power and energy storage system controls. 
% These findings suggest that system planners can implement network reconfiguration to accommodate higher load demands, then surely resulting in significant improvements in the HC. 
Additionally, we observe instances where PF control proves more effective than Q control and even ESS control in the IEEE 123-bus network. This highlights the potential for further study and adjustment of PF control to maximize HC. 
\vspace{-0.2 cm}

\subsection{HC estimation with both solar PV and wind turbine units}
\begin{figure}
    \centering
    \includegraphics[width=0.8\linewidth]{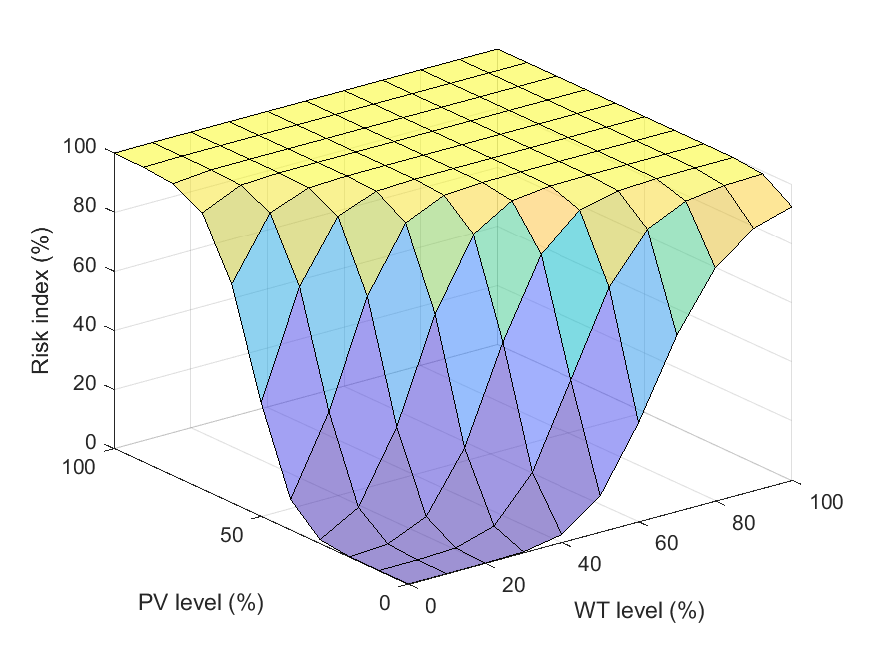}
    \vspace{-2mm}
    \caption{Variation in risk index with penetration levels of solar PV and wind turbine units}
    \vspace{-4mm}
    \label{fig:wind}
\end{figure}

We further utilized the proposed method to estimate the maximum renewable penetration when the network has both solar PV and wind turbine (WT) units. For a given level of solar PV and WT penetration, we randomly allocate these units to different buses using the method given in Appendix. We also considered the correlation between PV and WT units using copula function as shown in the Appendix. The results of variation in risk index with PV and wind turbine penetration is shown in Fig. \ref{fig:wind}. The data for this case study was taken from reference \cite{qin2013incorporating}. Further, we estimate the maximum WT penetration for a given penetration of solar PV and a given risk index levels. The results are shown in Table \ref{tab:my_label}. It can be observed that the amount of WT hosted by the network reduces if the penetration of PV increases.

\begin{table} 
    \centering
    \caption{GP-CC-HC estimation for PV and WT (\%) at risk index 5\%.}
    \label{tab:my_label}
    \begin{tabular}{c|cc}
        Risk level $\beta$ &  PV & WT\\
        \hline
     5\% & 10\% & 29.33\%  \\ 
         & 20\%  &  19.33\% \\
      & 30\% &   9.33\%\\ \hline
    \end{tabular}
    \vspace{-4mm}
\end{table}

\subsection{Notes on comparison with the existing method}
In this section, we aim to briefly compare our proposed methods with \cite{koirala2023chance}, where the authors employed a polynomial chaos expansion (PCE)-based chance-constrained probabilistic optimal power flow (PCE-CC-POPF) method. While both methods are effective for HC estimation in distribution networks, they exhibit differences in approach.

Firstly, the complexity of our proposed methods is relatively lower. We utilize a modest sample size, like $n_s=500$, for probabilistic load flow, then employ GPR and LoR models to predict over-voltage risk. Our chance-constrained optimization is solved with a straightforward objective function and univariate constraint. In contrast, the approach in \cite{koirala2023chance} is more complex, involving CC-POPF with multiple constraints and requiring simultaneous determination of coefficients for several PCE models.

Secondly, our proposed methods demonstrate superior computational efficiency. The GPR models are trained offline and can be applied to any nonlinear relationship with any distributions of correlated uncertain loads and renewable generations, and once the models are built, the online computation for predicting the HC values with different associated risks is fast and simple. In contrast, the general PCE-based POPF approach often encounters the ``\textit{curse of dimensionality}" and relies on the assumption of independent inputs \cite{ly2022scalable}. Thus, in practice, the degree of general multivariate PCEs is usually restricted to an order of 2 only. For highly nonlinear and correlated input scenarios, obtaining coefficients for general PCE models can be computationally intensive and time-consuming. Our methods, tested on IEEE 33-bus and 123-bus systems, complete the process of HC estimation in under 3.24 seconds (including training data generation), while the PCE-CC-POPF method in \cite{koirala2023chance} takes around 4.04 seconds for similar tasks, see Tab. III in \cite{koirala2023chance} for more time comparisons with other approaches. 
% Simulations were performed on a computer with an Intel Xeon processor 3.7 Ghz, 16 GB RAM, and using MATLAB 2019a.      

\subsection{Notes on pros and cons of the two proposed CC-HC estimation methods}
In this subsection, we provide a comparison of the pros and cons of the GP-based and logit-based HC estimation methods, as shown in Table \ref{tab:pros_cons}.

 \begin{table}[!ht]
     \centering
     \caption{The pros and cons of the GP-based and the logit-based chance-constrained HC estimation methods.}
   \resizebox{0.49\textwidth}{!}{  \label{tab:pros_cons}
     \begin{tabular}{|c|l|l|} \hline
         \textbf{Method} &  \textbf{Pros} & \textbf{Cons}\\ \hline
        \multirow{5}{*}{GP}  & 
        Captures non-linearity 	& Computational complexity \\
        & Non-parametric nature 	& Scalability issues \\ 
        & Uncertainty quantification & Hyperparameter tuning  \\
        & More robust to outliers &  \\
        & Smoothness assumptions & \\
         \hline
        \multirow{3}{*}{Logit} & Simplicity and interpretability	& Linear decision boundary \\
        & Computational efficiency	& Sensitivity to outliers \\
        &  & Limited to binary outcomes \\ \hline
     \end{tabular} }
     \vspace{-4mm}
\end{table}
 
The GP model is a non-parametric method, thus it does not assume a fixed form for the underlying function, allowing it to model complex, non-linear relationships effectively. Additionally, the GP-based method offers uncertainty quantification in predictions and is more robust to outliers, which is particularly useful in chance-constrained HC assessment where uncertainty is a critical factor. Furthermore, the GP assumes smoothness in the underlying function, making it well-suited for modelling continuous maximum voltage outputs, without the need to handle unbalanced data (e.g. zero-inflated binary outcomes) as in the logistic regression.  However, the GP approach comes with higher computational costs and scalability issues, especially for large datasets, as it involves inverting large covariance matrices. These issues can be solved by using sparse GP. In addition, the GP approach requires careful choosing kernel functions and tuning of their hyperparameters, which can be complex and time-consuming.

On the other hand, the logit-based method is simpler, more interpretable, and computationally efficient with large datasets. However, it may not capture complex relationships as effectively as GP, and it can be sensitive to outliers, owing to the assumption of a linear relationship between the predictor (i.e. the PV level) and the log-odds of the over-voltage event. Moreover, we need to apply zero-inflated models if the response variable contains an excess of zeros, such as in cases where violation events rarely occur.

\color{Black}
\vspace{-2mm}
\section{Conclusion} \label{sec5}
In this study, we proposed the chance-constrained PV HC estimation of distribution networks using Gaussian process regression and logistic regression. Our results showed that the GPR and LoR models can achieve high accuracy ranging from 90\%-93\% in predicting nodal over-voltage events in the IEEE 33 and 123-bus systems. Thus, they can be effectively applied to estimate the chance-constrained HC with different risk levels, especially the GPR can offer the mean values and confidence interval bounds for the HC estimation. Moreover, our proposed methods have simple forms and low computational costs with the total time spent of only a few seconds. In this work, we also confirmed that the use of Q, PF, and ESS controls can improve the HC of distribution networks. Our research has significant implications for the energy sector, aiding system operators in optimizing the integration of PV systems into distribution networks. %Future research should focus on the consideration of new uncertainties such as electric vehicle loads, local electricity markets and wind turbine distributed generations.
%Additionally, the impact of PV power curtailment on the network's HC also needs to be analyzed.

\vspace{-4mm}
\section*{\textbf{Appendix}}
\begin{figure}[tbp]
\vspace{-0.03 cm}
    \centering
\begin{tikzpicture}[font=\small,thick]
\node[draw,
    rounded rectangle,   draw=Blue,
    minimum width=2cm,
    minimum height=0.6cm] (block1) {Start, $k = 1$};
 
\node[draw,
    below=0.3 cm of block1,  draw=Blue,
    align=center,  
    minimum width=6cm,
    minimum height=0.5cm,
    text width=6cm,
] (block2) {Module 1: Obtain M representative load-generation profiles from historical data};
 
\node[draw,
    below=0.3 cm of block2, draw=Blue,
    align=center,  
    minimum width=6cm,
    minimum height=0.5cm,
    text width=6cm
] (block3) {Module 2: Generate N random location and size scenarios for PV installation};

\node[draw,
    below=0.3 cm of block3, draw=Blue,
    align=center,  
    minimum width=4cm,
    minimum height=0.5cm,
    text width=4cm
] (block4) {Module 3: Perform \textit{load flow} analysis for base case};

\node[draw,
    below=0.3 cm of block4, draw=Blue,
    align=center,  
    minimum width=5cm,
    minimum height=0.5cm,
    text width=5cm
] (block5) {Module 4: Obtain new inverter/ESS set-points considering their voltage control scheme and redo \textit{load flow}};

\node[draw, % Conditions test
    diamond, draw=Blue,
    below=0.3 cm of block5,  
    minimum width=2cm,
    minimum height=0cm,
    aspect=1.3, 
    inner sep=0] (block6) {Is $k = N_s$};
    
\node[draw,
    left=0.3 cm of block5,   draw=Blue,
    minimum height= 0.8cm, 
    minimum width=1.5cm,
    inner sep=0] (block7) {$ k = k + 1$};

\node[draw,
    rounded rectangle,   draw=Blue,
    below=0.7cm of block6,
    minimum width=2.5cm,
    minimum height=0.5cm,] (block8) {Stop};

\node[draw,
    right=0.3 cm of block4,   draw=Green,
    align=center,  
    minimum width=1cm,
    minimum height=0.5cm,
    text width=1cm] (block9) {Record voltage};

\node[draw,
    right=0.3 cm of block5,  draw=Green,
    align=center,  
    minimum width=1cm,
    minimum height=0.5cm,
    text width=1cm] (block10) {Record voltage};
 
\draw[-latex] (block1) edge (block2) % Arrows
    (block2) edge (block3) 
    (block3) edge (block4)
    (block4) edge (block5)
    (block5) edge (block6)
    (block4) edge (block9)
    (block5) edge (block10);

\draw[-latex] (block6) -| (block7)
    node[pos=0.25,fill=white,inner sep=0]{No};
    
\draw[-latex,line width=0.5mm] (block6) edge node[ pos=0.4,fill=white,inner sep=1pt]{Yes}(block8);
 
\draw[-latex] (block7) |- (block4);
 
\end{tikzpicture}
\vspace{-0.5 cm}
\caption{Flowchart of the data generation process.}
\vspace{-0.5 cm}
\label{fig:calc_rel}
\end{figure}
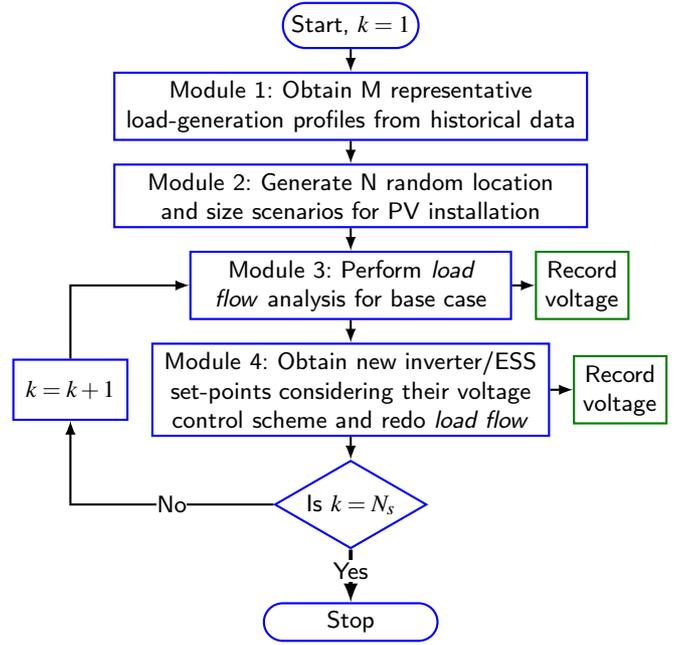

This appendix outlines the methodology for generating maximum voltage data for a given PV penetration level. Fig. \ref{fig:calc_rel} depicts the flowchart of the procedure. 
% The process starts with the generation of representative load-generation profiles using Module 1.  

%\vspace{-3mm}
\subsection*{Module 1: Representative load-generation profiles}
 
In this module, we sample uncertain load and PV output (normalized). 
% Further, we use copula theory to capture correlation between the variables.
% ~\cite{trivedi2007copula}.
These normalized values can be multiplied with base values to obtain a random scenario for analysis. 
Let $F_1$ and $F_2$ be CDFs of the normalized load $P_{dn}$ and normalized PV generation $P_{gn}$. Then by Sklar's theorem in the copula theory \cite{trivedi2007copula}, there exists a unique copula function $C$ capturing the dependence structure/correlation between the variable $P_{dn}$ and $P_{gn}$ such that the joint CDF defined by:
\begin{align}
H(p_1,p_2) & := \mathbb{P}\big(P_{dn} \le p_1, P_{gn}\le p_2\big) \nonumber \\
& = C\big(F_1(p_1),F_2(p_2)\big). 
\end{align}
It is well-known that if $P_{dn}$ and $P_{gn}$ are independent, then their copula is $C(u_1,u_2)=u_1u_2,$ and we have the joint CDF:  $H(p_1,p_2) = F_1(p_1)F_2(p_2)$ \cite{trivedi2007copula}. %\cite{ly2019determining_product,ly2019determining_ratios}. 
% Otherwise, there are many other copula families such as Gaussian, Student-t, Clayton, Gumbel, Frank, Joe, etc., that can be used to model the dependence structure between the normalized load  $P_{dn}$ and normalized PV $P_{gn}$.
For the sake of simplicity, we shall use a Gaussian copula  $C_\rho(u_1,u_2)$, where $\rho$ is the copula parameter (Pearson's correlation coefficient). Also, we take the marginal CDFs $F_1$ and $F_2$ as normal $\mathcal{N}(0.5, 0.025^2)$ and $\text{Beta}(15, 6)$ distributions, respectively. Now, we can generate the representative load and generation profiles as follows:
\begin{itemize}
    \item[(i)] Generate $M$ random samples using estimated copula: $$(u_{1}^{j},u_{2}^{j}) \sim C_\rho,\, j=1,\ldots, M.$$ 
    % This can be done with the help of the package \textit{copula} in R, or Patton's copula toolbox in MATLAB. 
    \item [(ii)] Get normalized load and PV generation profiles by:
    $$ P_{dn}^j=F_{1}^{-1}\big(u_{1}^j\big), \quad  P_{gn}^j=F_{2}^{-1}\big(u_{2}^j\big), \, j=1,\ldots, M.$$
    \item [(iii)] Multiply normalized values with base values. 
    % to obtain random scenarios for analysis.
\end{itemize}

\noindent In this work, we have not included spatial correlation explicitly. Since the size of a distribution network is rather small, it seems reasonable to assume uniform solar conditions.

%\vspace{-3mm}
\subsection*{Module 2: Generating location and size scenarios}
In this module, we create $L$ location-size scenarios of the PV unit installations. Let $PV^{loc}$ be the set of candidate buses at which PV units can be installed with $N^{pv}$ being the set size. For each scenario $i$ in the set $L$, firstly, a vector $u$ is initialized.
% having size $N^b \times 1$ and all elements equal to 0 is initialized. Here $N^b$ is the total number of buses in the network. T
This vector represents the size of the PV unit installed at each bus. Next, the total number of buses having PV units (denoted as $K$) is obtained randomly from a discrete uniform distribution. This is followed by randomly selecting $K$ unique integers from the set $loc$ to formulate the set $PV^{loc}$ of random location scenarios. Finally, for each bus in the set $PV^{loc}$, the PV is sized by generating a random number between 0 and $\varsigma$ using the uniform distribution. 
% The value $\varsigma$ depends on the network operator's PV policy.
We have limited the PV installed at each bus to 1.5 times the peak load for realistic analysis. 
%%% Finally, the vector $u$ is stored in matrix $\Psi^{size}$ and the total PV size is stored in vector $\Psi^n$. 
% The detailed algorithm for the process is given in Algorithm \ref{Alg:loc-size}.

%\vspace{-3mm}

\subsection*{Module 3: Base case \textit{load flow}}
After obtaining location-size scenarios from Module 2, load flow analysis is performed. For each scenario, M load flows are performed corresponding to M representative load-generation profiles. The PV output at each bus is derived by multiplying the PV size at that bus by the normalized generation values from Module 1. This PV output is then subtracted from the load to determine the net load.
% The load considered here is also obtained by multiplying the representative load profile (normalized) by the peak active and reactive power demand at each bus. 
Further, the reactive power from the PV inverter is considered to be 0 in this module. 
% The maximum voltage obtained for each scenario is stored to be later used in learning the relationship between risk and PV installed. 

%\vspace{-3mm}
\subsection*{ Module 4: \textit{Load flow} considering PV inverter/ESS control}
In this module, we consider the case where the PV inverters are equipped with a localized voltage control algorithm. The voltages obtained from base-case load flow (i.e. Module 3) are the voltages sampled by the PV inverters. Based on this measured voltage, a new set-point is obtained
% for the active and reactive power output from the PV inverter 
based on the control algorithm.
In this work, we have considered Volt-Var (Q) control and power factor (PF) control \cite{ibrahim2022benchmark}.
% to analyze their effects on the network's HC 
% \cite{ibrahim2022benchmark}
The Volt-Var technique to control voltages is as follows:
\begin{align}\label{eq:qc1}
\begin{split}
    &Q_{pv,i} = \begin{cases}
			Q_{max,i}, & \text{if $V_i \leq v_1$},\\
            \frac{V_i - v_2}{v_1 - v_2}\, Q_{max,i}, & \text{if $v_1 <  V_i \leq v_2$}, \\
            0, & \text{if $v_2 < V_i < v_3$}, \\
            -\frac{V_i - v_{3}}{v_{4} - v_{3}}\, Q_{max,i}, & \text{if $v_3 \leq  V_i < v_4$}, \\
            -Q_{max,i}, & \text{if $V_i \geq v_4$},
		 \end{cases}\\
   &Q_{max,i} = \sqrt{S_{max,i}^2 - P_{pv,i}^2},
   \end{split}
\end{align}
where $P_{pv,i}$ is the active power output of the PV unit and $S_{max,i}$ is the inverter rating. In this work, we have assumed an inverter rating equal to the PV unit size. It can be observed that there is no curtailment of active power in this method. The values of $(v_1, ... , v_4)$ are (0.95,0.97,1.03,1.05) per unit.

Next, we consider the case of the adaptive power factor method. In this method, the PV inverter changes its power factor according to the voltage value as follows:
\begin{align}\label{eq:qc2}
\begin{split}
    &PF_{pv,i} = \begin{cases}
			1, & \text{if $V_i \leq v_3$},\\
            c_1 + \frac{1 - c_1}{v_3 - v_4}\, (V_i - v_4), & \text{if $v_3 <  V_i < v_4$}, \\
            c_1, & \text{if $V_i \geq v_4 $}, 
		 \end{cases}
   \end{split}
\end{align}
where $c_1$ is chosen as 0.95 in this work.

\noindent Lastly, we define the voltage control using ESS as follows:
\begin{align}\label{eq:ess}
\begin{split}
    &P_{ess,i} = \begin{cases}
			P_{max,i}, & \text{if $V_i \leq v_1$},\\
            \frac{V_i - v_2}{v_1 - v_2}\, P_{max,i}, & \text{if $v_1 <  V_i \leq v_2$}, \\
            0, & \text{if $v_2 < V_i < v_3$}, \\
            -\frac{V_i - v_{3}}{v_{4} - v_{3}}\, P_{max,i}, & \text{if $v_3 \leq  V_i < v_4$}, \\
            -P_{max,i}, & \text{if $V_i \geq v_4$},
		 \end{cases}\\
   \end{split}
\end{align}
where $P_{max,i}$ is the rated power output of $i^{th}$ ESS. We considered 6 ESS with a rating of 150 kW each and located at bus $[$50, 70, 72, 80, 105, 110$]$ in the 123-bus network. 
A negative sign signifies ESS charging.
% We have also taken an assumption that the control action does not violate the ESS state of charge limits.

Finally, similar to Module 3, load flow analysis is performed after adjusting the PV unit's output into active/reactive power demand at each bus. The control procedure iterates until convergence, defined as the maximum voltage magnitude difference between iterations being less than $\epsilon_v$ (0.005). The detailed procedure of voltage control in Module 4 is provided in the algorithm (Algorithm \ref{Alg:qc1}).
% Finally, the voltage maximum for each scenario is stored to be later used in learning the relationship between risk and PV installed. 
 
\begin{algorithm}[H]
\begin{algorithmic}[1]
\STATE \textbf{Input: $V, \epsilon_v$}
\STATE \textbf{Initialize: $\Delta V = \mathbf{0}, err = 1,$}
\STATE \textbf{Output: $P_{pv},Q_{pv}$}
\WHILE{ $err > \epsilon_v $}
\FOR{each bus}
        \STATE Obtain new PV inverter/ ESS set-points using \eqref{eq:qc1} / \eqref{eq:qc2} / \eqref{eq:ess}. 
\ENDFOR
\STATE Using new inverter/ ESS set-points, obtain bus voltages ($V_n$) using \textit{load flow} analysis.
\STATE $err = \max(|V - V_n|)$
\STATE $V = V_n$
\ENDWHILE
\end{algorithmic}
\caption{ Module 4: Obtain new setpoints for PV inverter/ESS and re-perform load flow}
\label{Alg:qc1}
\end{algorithm}

\bibliographystyle{abbrv}
%\bibliography{HC_ref}

\end{document}